# *Ab initio* calculation of the dynamical properties of PPP and PPV

R. L. de Sousa and H. W. Leite Alves[†]

*Departamento de Ciências Naturais, Universidade Federal de São João del Rei, Caixa Postal 110, 36300-000 São João del Rei-MG, Brazil*

In this work, we have calculated the vibrational modes and frequencies of the crystalline PPP (in both the Pbam and Pnnm symmetries) and PPV (in the $P2_1/c$ symmetry). Our results are in good agreement with the available experimental data. Also, we have calculated the temperature dependence of their specific heats at constant volume, and of their vibrational entropies. Based on our results, at high temperatures, the PPP is more stable in the Pnnm structure than in the Pbam one.

## I. Introduction

The interest in the electronic and optical properties of semiconducting π-conjugated polymers, such as the poly-paraphenylene (PPP) and the poly-paraphenylene-vinylene (PPV) has considerably increased due to the necessity on the improvement of the efficiency of the new polymer-based optoelectronic devices, such as LEDs and flat plane displays. This efficiency is determined by the fraction of injected electrons and holes that recombine to form emissive spin-singlet states from the excitons rather than non-emissive spin-triplet states [1].

In conventional semiconductors such as Si, the Wannier excitons are well defined as weakly bound electron-hole pairs. In organic crystals such as anthracene, on the contrary, the Frenkel exciton is essentially confined to a single molecule leading to high binding energies. However, in the π-conjugated polymers, these excitations were in the intermediate case, current subject of debate in the academic community, because it is very hard to describe the vibronic states in the theoretical treatment of the excitations of these systems [2]. Besides that, accurate experimental data for the phonon modes on these materials available in the literature are very scarce: only the strongest and the medium active modes were assigned by infrared and Raman experiments [3-8]. Moreover, on the theoretical side, only simplified models, based on the linear chain approaches, were used to explain the observed experimental data [9-11].

In order to supply the missing information on the vibrational properties of PPP, in both the Pbam and Pnnm structures, and of PPV in the $P2_1/c$ crystalline symmetry, we have calculated, in this work, by using the Density Functional Theory (DFT) within the Local Density Approximation (LDA), plane-wave description of the wave functions and the pseudopotential method (ABINIT code) [12], their frequencies and the correspondent vibrational modes. In our calculations, we have used the Troullier-Martins pseudopotentials (evaluated by the fhi98PP code [13]), and the phonons were obtained by means of the Density-Functional Perturbation Theory. Details about the structural properties of these systems, as well as the vibrational modes, obtained by simplified models, are described in our previous work [14, 15].

It is well known that while bonding among the atoms within the polymer chains mainly is of covalent character, interactions between the polymer chains are governed by weak van der Waals (vdW) forces. In this case, the common LDA and GGA fail to describe this weak nonlocal force, mainly in noble gases [16]. However, correlation functionals, like LDA and GGA, contract the density towards the bond and valence region thus taking negative charge out of the vdW region. Moreover, for large interchain separation $d$, the vdW interaction energy between chains behaves as $d^{-4}$, while LDA, or GGA, predicts an exponential falloff. In polymers, as also observed in fullerenes, it behaves as $d^{-6}$ [17]. So, the LDA is a good choice to describe these systems, as already shown in previous works for both PPP and PPV [18, 19].

## II. PPP phonon dispersion

In Figs. 1 and 2, we show the phonon dispersions and densities of states (DOS) for the PPP in the Pbam and

---

[†] Corresponding author: e-mail: hwlalves@ufsj.edu.br, tel.: +55 32 33.79.24.89, fax: +55 32 33.79.24.83

Pnnm crystalline symmetries, respectively, at selected points of the Brillouin zone.

The main difference among these two structures (Pbam and Pnnm) is a shift of one molecular chain, related to the other, along the chain direction by an amount of *c/2*. From our calculated structural parameters, there is almost no remarkable difference between these two modifications [14]. Moreover, despite the fact that our results show that the Pnnm is 0.54 meV more stable than the Pbam, from the accuracy of our calculations, 1 meV, we can only infer that both structures can coexist at low temperatures.

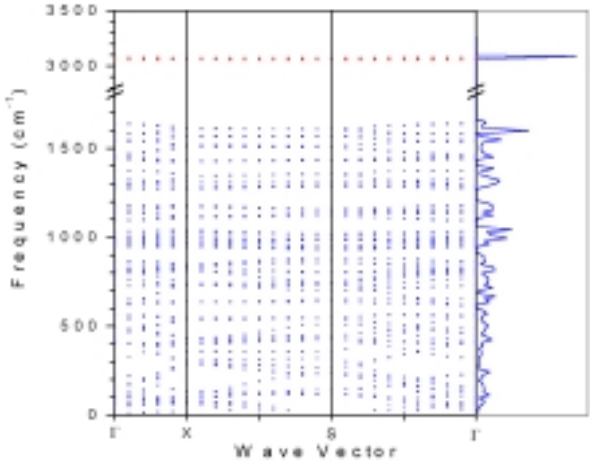

Fig. 1. Calculated phonon dispersion and density of states for the PPP in the Pbam crystalline structure at selected points of the Brillouin zone. The Γ-X direction is parallel to the polymer chain, while the X-S direction is perpendicular to the chain.

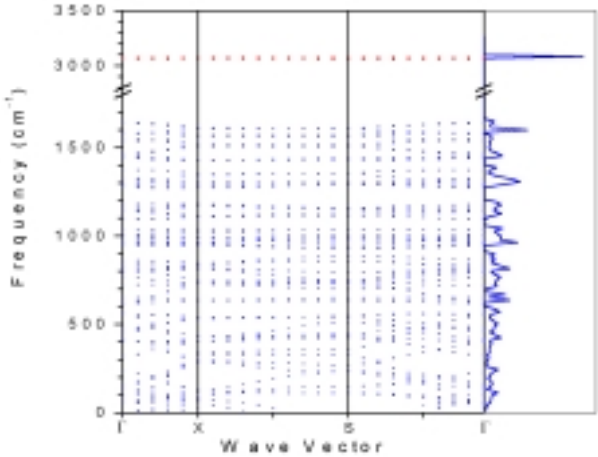

Fig. 2. Same as Fig. 1 for PPP in the Pnnm crystalline structure.

From Figs. 1 and 2, if we look only for the calculated phonon dispersions, we have also detected that there is no remarkable difference between the Pbam and Pnnm structures. However, after a careful analysis of both evaluated DOS, we have noted that there are shifts in some peaks as follows: i) the peak at 1500 cm$^{-1}$ in the Pbam DOS appears at 1300 cm$^{-1}$ in the Pnnm one; ii) the peak at 1050 cm$^{-1}$ in the Pbam DOS appears at 800 cm$^{-1}$ in the Pnnm one; and finally, iii) there is a rearrangement of the low frequency modes below 100 cm$^{-1}$ between both calculated DOS. Although the differences between the two evaluated DOS are relatively small, they should be sufficient for an interesting difference in their derived thermodynamic properties.

Concerning the vibrational modes, the narrow flat band, located between 3054 and 3077 cm$^{-1}$ in both dispersions, corresponds to C-H stretching and bending vibrations. We have noted that the modes located at the frequency region from 1035 to 1642 cm$^{-1}$ are related to the C-C stretching and bending vibrations. Also, the modes located at the frequency region from 726 to 1035 cm$^{-1}$ are C-C stretching vibrations only, while the modes with frequency lower than 726 cm$^{-1}$ are C-C bending vibrations.

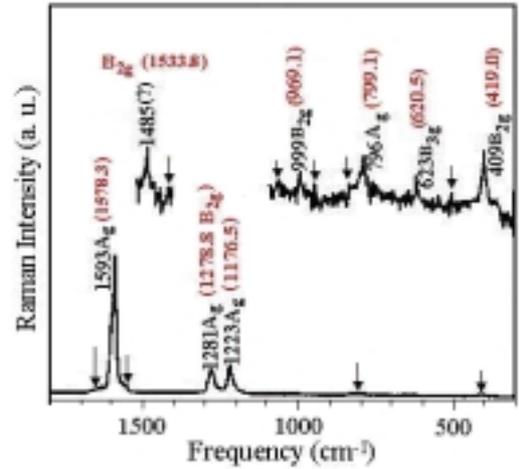

Fig. 3. Experimental Raman spectra for PPP (taken from Ref. 5) compared with our obtained results (the numbers in the parenthesis). The arrows indicate our assignment of some unidentified Raman peaks based on our data. The upper curves are an amplification of some areas of the spectra depicted below.

Both the symmetry characters and the frequency values of our calculated vibrational modes are in good agreement with both Raman and infrared results [3-5]. In Fig. 3, we show our calculated phonon frequencies (the numbers in the parenthesis) compared with the Raman spectra obtained by Furukawa *et al.* [5]. From Fig. 3, we have identified the symmetry character of the 1485 cm$^{-1}$ mode, $B_{2g}$. However, the symmetry character of the experimental 1281 cm$^{-1}$, $A_g$, is in contradiction with our calculated $B_{2g}$, instead. Moreover, based on our obtained data, we have assigned some unidentified peaks, depicted as arrows in Fig. 3, as follow: a $B_{1g}$ mode at 1637 cm$^{-1}$, a $B_{3g}$ mode at 1535 cm$^{-1}$, an $A_g$ mode at 1328 cm$^{-1}$, an $A_g \oplus B_{1g}$ mode at 1007 cm$^{-1}$, a $B_{2g}$ mode at 946 cm$^{-1}$, a $B_{1g}$ mode at 843 cm$^{-1}$, and a $B_{2g}$ mode at 488 cm$^{-1}$. A complete description of our obtained results will be published soon in another publication.

## III. PPV phonon dispersion

In Fig. 4, we show the phonon dispersion and DOS for the PPV in the P2$_1$/c crystalline symmetry, at selected points of the Brillouin zone.

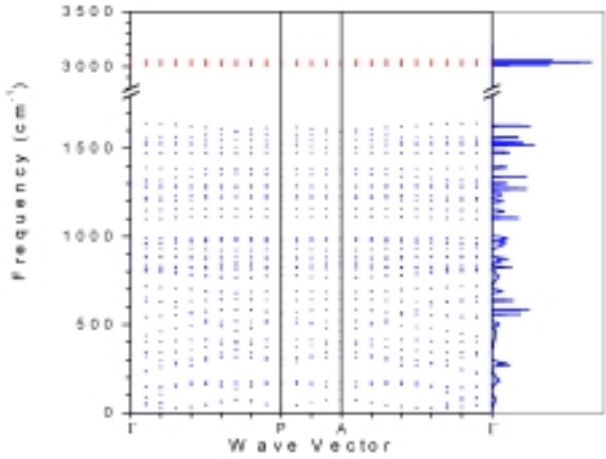

Fig. 4. Same as Fig. 1 for PPV in the P2$_1$/c crystalline structure. Here, the Γ-P direction is parallel to the polymer chain and the P-A direction is perpendicular to the chain.

From Fig. 4, the narrow flat bands, located between 3010 and 3060 cm$^{-1}$, corresponds to C-H bond stretching and bending vibrations. However, these flat bands consist in three distinct regions: the first two, one at 3010-3025 cm$^{-1}$ and the other at 3035-3042 cm$^{-1}$, are concerned to vinyl C-H vibrations; while the last, at 3050-3060 cm$^{-1}$, is related to the phenyl C-H vibrations, and it is around 11-21 cm$^{-1}$ below the calculated frequency values of these modes on PPP.

Also from Fig.4, we have noted that the modes located at the frequency region from 1520 to 1645 cm$^{-1}$ are related to the vinyl C-C stretching vibrations. The vinyl C-C bending modes combined with phenyl C-C, stretching or bending, vibrations are located at the frequency region from 804 to 1520 cm$^{-1}$. Finally, the modes with frequency lower than 750 cm$^{-1}$ are torsions of both the phenyl and vinyl radicals.

Moreover, both the symmetry characters and the frequency values of our calculated vibrational modes are in good agreement with both Raman and infrared results [7, 8]. We show, in Fig. 5, our calculated phonon frequencies (the numbers in the parenthesis) compared with the experimental Raman results of Orion et al. [7]. From this comparison, besides the agreement of our symmetry character assignment of the known experimental modes, we have identified some unidentified ones, depicted as arrows in Fig. 5: an $A_g$ mode at 961 cm$^{-1}$, an $A_g$ mode at 1299 cm$^{-1}$, and an $A_g$ mode at 1518 cm$^{-1}$.

Also, our results agree with recent site-selective luminescence spectroscopy results obtained by Borges et al. [20]. In their experimental results, the PPV spectra can be well described by three effective vibrational modes at 330, 1160 and 1550 cm$^{-1}$. From our calculated results, these modes are an $A_g$ one (at 329 cm$^{-1}$), a $B_g$ one (at 1150 cm$^{-1}$) and an $A_g$ one (at 1556 cm$^{-1}$). While the last mode is related only to the C-C stretching in the vinyl, the other two modes involve the entire chain. This is the reason that the observed Huang-Rhys factors of the first two modes have an interesting variation with the temperature increasing. If the Huang-Rhys factors are related to the conjugation length, the modes, which involve the entire chain, are the most affected ones [20]. Further details of our obtained results will be published soon in another publication.

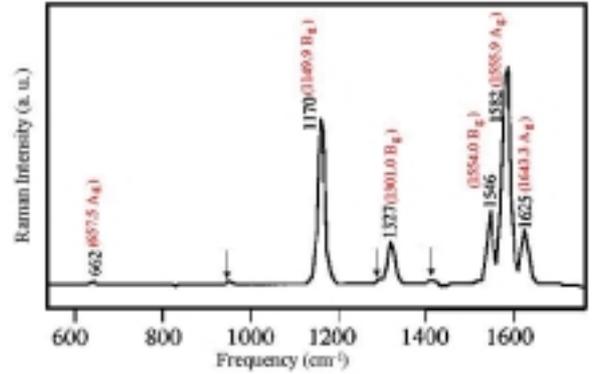

Fig. 5. Same as Fig. 3 for PPV (the experimental data were taken from Ref. 7).

## IV. Thermodynamic properties

We also have calculated some thermodynamic properties such as the constant volume specific heat and the vibrational entropy. The temperature dependence of the specific heat for PPP and PPV was evaluated as follows,

$$C_V = 3rNk_B \int_0^{\omega_M} \left(\frac{\hbar\omega}{2k_BT}\right)^2 \cos\sec h^2\left(\frac{\hbar\omega}{2k_BT}\right) D(\omega)d\omega \quad,$$

where $N$ is the number of the cells, $r$ is the number of the atoms of the basis, $k_B$ is the Boltzmann constant and $D(\omega)$ is the density of phonon states. The vibrational entropy was evaluated as

$$S_{Vib} = 3rNk_B \times$$
$$\times \int_0^{\omega_M} \left[\frac{\hbar\omega}{2k_BT}\coth\left(\frac{\hbar\omega}{2k_BT}\right) - \ln\left\{2\sinh\left(\frac{\hbar\omega}{2k_BT}\right)\right\}\right] D(\omega)d\omega \quad.$$

In Fig. 6 and 7, we show our evaluated temperature dependence of the specific heat and vibrational entropy, respectively, for PPP and PPV. From Fig. 6 we have found that the PPV has lower heat capacity than PPP. From Fig. 7, we have noted that the Pnnm modification is more stable than the Pbam structure at high temperatures, once its entropy is slightly the greatest.

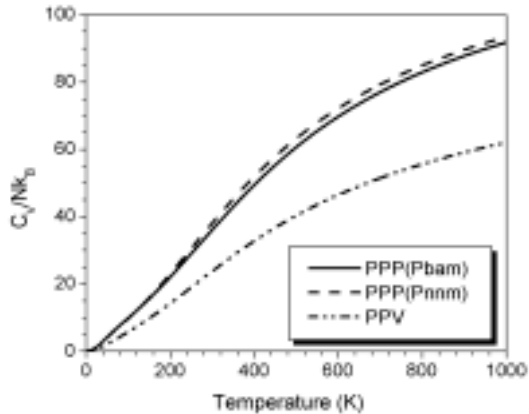

Fig. 6. Calculated temperature dependence of the specific heat for both PPP and PPV.

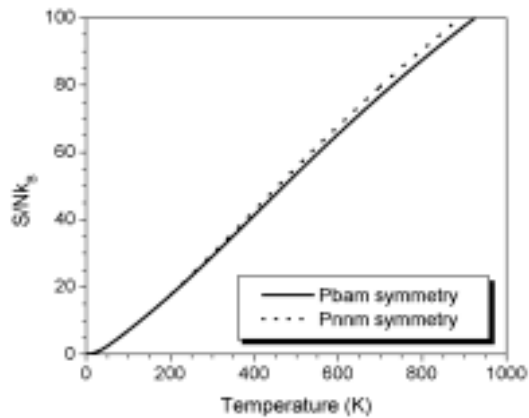

Fig. 7. Calculated temperature dependence of the entropy for the PPP.

## V. Conclusions

In summary, we have presented our *ab initio* results for the vibrational modes of the PPP (in both the Pbam and Pnnm symmetries) and of PPV. Our results are in good agreement with the available experimental data and, we have assigned some unidentified Raman peaks which have weak activity. Also, we have shown some calculated thermodynamic properties, which show that the PPP is more stable in the Pnnm structure at high temperatures and, it has higher heat capacity than the PPV. The inclusion of interchain effects to obtain the vibrational modes of these polymers is, based on our results, very important, as we have already mentioned recently [15]. We hope that our results give guidelines for future experiments on this subject.

## Acknowledgements

R. L. de Sousa is indebted to the Brazilian Agency, MEC-CAPES, for her fellowship during the development of this research. This work was also supported by the MCT/CNPq/PRONEX project No. 662105/98, Brazil.